# InGaN/GaN Multi-Quantum-Well and Light-Emitting Diode Based on V-pit-Shaped GaN Grown on Patterned Sapphire Substrate

Lai Wang[*], Xiao Meng, Di Yang, Zilan Wang, Zhibiao Hao, Yi Luo, Changzheng Sun, Yanjun Han, Bing Xiong, Jian Wang and Hongtao Li

*State Key Laboratory on Integrated Optoelectronics / Tsinghua National Laboratory for Information Science and Technology, Department of Electronic Engineering, Tsinghua University, Beijing 100084, P. R. China*

---

[*] Corresponding author. Tel: +86-10-62782734; Fax: +86-10-62784900

*E-mail address:* wanglai@tsinghua.edu.cn (L. Wang)

**Abstract**

V-pit-defects in GaN-based light-emitting diodes induced by dislocations are considered beneficial to electroluminescence because they relax the strain in InGaN quantum wells and also enhance the hole lateral injection through sidewall of V-pits. In this paper, regularly arranged V-pits are formed on c-plane GaN grown by metal organic vapor phase epitaxy on conventional c-plane cone-patterned sapphire substrates. The size of V-pits and area of flat GaN can be adjusted by changing growth temperature. Five pairs of InGaN/GaN multi-quantum-well and also a light-emitting diode structure are grown on this V-pit-shaped GaN. Two peaks around 410 nm and 450 nm appearing in both photoluminescence and cathodeluminescence spectra are from the semipolar InGaN/GaN multi-quantum-well on sidewalls of V-pits and c-plane InGaN/GaN multi-quantum-well, respectively. In addition, dense bright spots can be observed on the surface of light-emitting diode when it works under small injection current, which are believed owing to the enhanced hole injection around V-pits.

## 1. Introduction

GaN-based light-emitting diodes (LEDs) have achieved great success on display and solid state lighting applications in recent years.[1] As the active region of LEDs, InGaN/GaN multi-quantum-well (MQW) plays an important role on performance of LEDs. By transmission electron microscopy characterization, it is found that in InGaN/GaN MQW there exists a type of frequently observed defects, called as V-pit defects.[2-7] A V-pit defect is usually formed by a threading dislocation in GaN bulk layer when it penetrates into InGaN/GaN MQW active region, which is regard as a way of strain relaxation in InGaN in virtue of dislocation. Since the dislocations density in hetero-epitaxial GaN is very considerable, the influence of V-pit defects cannot be ignored. A. Hangleiter, *et al*. proposed in 2005 that V-pit defect is a possible origin why internal quantum efficiency of InGaN MQW is insensitive to dislocation density.[8] As the InGaN MQW on sidewalls of V-pit defect is thinner than the planar MQW, the energy level of the former will be higher than that of the latter. As a result, there will form a blocking barrier around the V-pit defect to avoid carrier diffusion to dislocation. In addition, L. C. Le *et al*. also proved that V-pit defects were a type of carrier leakage channels in LEDs under certain cases.[9] However, it was found recently that V-pit defects were also possibly beneficial for LEDs, because the hole can be laterally injected into InGaN MQW through sidewalls of V-pits.[10-14] Thus, the hole concentrations distribution uniformity in different quantum wells can be improved. On the other hand, as we know, the strain in InGaN QW will be accumulated as the number of QW increasing. Since V-pit defects are the results of strain relaxation, the number of QWs can be also increased with the help of V-pits, which will enhance the carrier capture and improve the LED performance under high injection. Due to these advantages, V-pit defects have been utilized in most current commercialized LEDs, wherein an InGaN/GaN superlattice structure is introduced beneath the InGaN/GaN MQW active region. This supperlattice structure can lead to formation of V-pit defects previously. By adjusting the indium composition or cycle number of superlattice, the average density and the size of V-pit defects in MQW can be well controlled. However, since V-pits are originated from dislocations, their positions are still random and the uniformity is also not as good as expected. In addition, InGaN/GaN superlattice will also prolong the epitaxial growth time of LED, because commonly its growth rate is around 1.5-2 nm/minute and its total thickness exceeds 100 nm.

Recently, we propose and demonstrate to inductively grow GaN with regularly arranged V-pit defects on widely-used patterned sapphire substrates (PSS). [15] In the present work, we continue to grow InGaN/GaN MQW and LEDs on this V-pits-shaped GaN and observe the electroluminescence enhancement around the V-pits.

## 2. Experiments

The metal organic vapor phase epitaxy is carried out in an Aixtron 2000HT system. By decreasing the growth temperature of GaN to around 950 °C, V-pits-shaped GaN is obtained on 2-inch PSS, wherein the bottom of V-pit is just over the top of sapphire cone. The detailed growth conditions are similar to those reported in Ref. 15. Subsequently, an MQW sample consisting of five-pairs of blue InGaN/GaN MQW and an LED sample containing such MQW are continuously grown on this V-pits-shaped GaN respectively. For the MQW sample, the nominal thickness of InGaN and GaN are 3 nm and 10 nm, respectively. For the LED sample, Si dopants are introduced in V-pits-shaped GaN, while a 50-nm p-$Al_{0.1}Ga_{0.9}N$ and a 200-nm p-GaN are grown on MQW in succession. After growth, the MQW sample is characterized by scanning electron microscopy (SEM), cathodeluminescence (CL) and photoluminescence (PL). For the LED sample, the ITO transparent electrode is deposited on the surface for electroluminescence (EL) fast test. Owing to the rough surface, n-GaN etching is omitted and the n-type electrode is replaced by an indium bump contact at the edge of wafer.

## 3. Results and Discussion

Figure 1(a) shows the surface morphology of the MQW sample. Regular V-pits arrange hexagonally, following the patterns of substrate strictly. The details about V-pit are shown in Fig. 1(b). It can be seen that there are totally 12 sections inside the V-pits, wherein the dark and bright ones correspond to $(11\bar{2}2)$ and $(1\bar{1}01)$ facets, respectively. These two types of semipolar facets are the most common ones those obtained by selective area growth (SAG) on GaN. [16-20] From Fig. 1(c), it is clear to see that the V-pit is just over the top of sapphire cone. And the tilt angle of semipolar planes is around 60°, which coincides to the facets mention above.

The MQW sample is characterized by CL and PL, as shown in Fig. 2. Fig. 2(a) is the SEM photo of the surface. The flat region corresponds to c-plane MQW, while the V-pit region corresponds to semipolar MQW. The CL spectra of the whole region, flat region and V-pit region are measured respectively and plotted in Fig. 2(b). The voltage of the electron beam is chosen to 5 kV. The intensity is normalized to the highest peak in each spectrum. Besides the GaN peak around 365 nm, there are two peaks originated from InGaN/GaN MQW within blue wavelength range. The one around 407 nm is from semipolar MQW inside the V-pit, while the other one around 457 nm is from c-plane MQW at flat region. This result is owing to two reasons: first, the semipolar MQW has the lower polarization field and hence the weaker quantum confinement Stark effect (QCSE), leading to shorter wavelength; second, the thickness of InGaN quantum well inside the V-pit is much thinner than that of c-plane InGaN quantum well due to a lower growth rate, which has been verified in previous publications. The CL mapping photos of 407 and 457 nm emission are shown in Fig. 2(c) and (d), respectively. The bright patterns in these photo can further confirm the origins of these two peaks are from V-pits and c-plane region, respectively. However, from Fig. 2(c), it is found that the bright patterns only appear in $(11\bar{2}2)$ facets, while $(1\bar{1}01)$ facets do not contribute to emission. This phenomenon seems not the same with the previous report, [17] and further study is desired to clarify the origin. The excitation power dependent PL measurement is carried out at room temperature using a 20-mW 325-nm He-Cd laser as the excitation source. The excitation power is decayed by attenuators to 2, 0.2, and 0.02 mW, respectively. The PL spectra under different excitation power are plotted in Fig. 2(e). Similar to CL spectrum, two peaks around 407 and 450 nm are observed, corresponding to semipolar MQW and c-plane MQW respectively. From comparison of spectra under different excitation power, it can be see that the peak wavelengths almost keep constant. In polar InGaN MQW, the PL wavelength usually exhibits blue-shift as excitation power increasing, which is a reflection of carrier screening to polarization induced QCSE. [21,22] Therefore, the observed unchanging of peak wavelengths indicates that the polarization in either semipolar MQW or c-plane MQW is reduced. The reason for the former is obvious, since $(11\bar{2}2)$ plane is nearly a polarization-free plane. For the latter, it is considered that V-pits defects reduce the strain in c-plane InGaN MQW, thus the piezoelectric polarization in c-plane MQW is also weakened.

The LED sample based on the MQW grown on V-pit-shaped GaN is also characterized by SEM and CL mapping. Figure 3(a) shows the surface morphology of the LED sample. The V-pits become more like a hexagon in the LED sample. From the detailed picture of V-pit shown in Fig. 3(b), it is found that this change is induced by the diminishment of $(11\bar{2}2)$ facets and expansion of $(1\bar{1}01)$ facets. This competition between two facets might be determined by growth parameters. The CL spectra of the whole region, flat region and V-pit region are measured respectively and plotted in Fig. 3(c). Similar to Fig. 2(b), two peaks of around 417 nm and 457 nm are observed. However, unlike the MQW sample, the c-plane MQW peak dominates the CL spectrum in the LED sample. The CL mapping at these two peak wavelengths are also measured and shown in Fig. 3(d) and (e), respectively. The patterns in Fig. 3(e) clearly show that the 457 nm emission is from c-plane MQW, while the patterns of 417 nm emission in Fig. 3(d) becomes slightly fuzzy. However, it can be still seen indistinctly that the bright regions are $(11\bar{2}2)$ facets. Thus, the diminished area of $(11\bar{2}2)$ facets is reasonable to explain why the emission from semipolar MQW inside V-pits becomes weaker in the LED sample. According to this result, if the diminishment of $(11\bar{2}2)$ facets are induced by p-AlGaN and p-GaN growth, it is unlikely that their emission intensity will be reduced significantly. Thus, it can be speculated that the diminishment of $(11\bar{2}2)$ facets happens more probably during the n-GaN or MQW growth. In the LED sample, Si-doping is introduced in n-GaN and the barriers of MQW. Si dopants are believe a surfactant in MOVPE of GaN, so this would result in the competition between $(11\bar{2}2)$ and $(1\bar{1}01)$ facets. Whatever, the dominant luminescence is from the c-plane MQW, indicating the LED sample is still more like the current normal LED.

The EL emission photos of the LED sample under fast test are listed in Fig. 4. Dense bright spots are observed under small injection current. These bright spots are commonly observed in InGaN-based LEDs. They are believed related to enhanced hole lateral injection through the sidewalls of V-pit defects, which can improve the distribution uniformity of hole in MQW.[14] In previous work, the V-pits are induced by threading dislocation, while in the present work, they are inductively grown on PSS. The density of bright spots is also higher than the normal LED. However, since the V-pit defects occupy the surface area, the area of c-plane MQW is much smaller than the normal LED. This means the size of V-pit should be optimized in further study to achieve the stronger EL intensity.

## 4. Conclusions

Blue InGaN/GaN MQW and LED based on V-pit-shaped GaN are grown on PSS by MOVPE. Two peaks appearing in CL and PL measurements are attributed from semipolar MQW inside the

V-pits and c-plane MQW on flat region, respectively. The semipolar MQW emission is mainly from $(11\bar{2}2)$ facets. In LED sample, c-plane MQW emission dominates the spectrum and semipolar MQW emission is weakened due to the diminishment of $(11\bar{2}2)$ facets. In fast test of LED sample, dense bright spots are observed on the surface, which are considered due to the enhanced hole lateral injection through V-pit sidewalls.

**Acknowledgements**

The authors are thankful for the supports of National Key Research and Development Plan (Grant No. 2016YFB0400102), National Basic Research Program of China (Grant Nos. 2012CB3155605, 2013CB632804, 2014CB340002 and 2015CB351900), the National Natural Science Foundation of China (Grant Nos. 61574082, 61210014, 61321004, 61307024, and 51561165012), the High Technology Research and Development Program of China (Grant No. 2015AA017101), Tsinghua University Initiative Scientific Research Program (Grant No. 2013023Z09N, 2015THZ02-3), the Open Fund of the State Key Laboratory on Integrated Optoelectronics (Grant No. IOSKL2015KF10), the CAEP Microsystem and THz Science and Technology Foundation (Grant No. CAEPMT201505) and S&T Challenging Project.

**Figure captions**

Fig. 1 SEM photos of the MQW sample. (a) Surface morphology of regularly arranged V-pits; (b) Details on sidewalls of a V-pit. The Dark and bright regions correspond to $(11\bar{2}2)$ and $(1\bar{1}01)$ facets, respectively; (c) Cross-sectional SEM of a V-pit. The tilt angle of the sidewall is around 60°.

Fig. 2 CL and PL properties of the MQW sample. (a) SEM photo of the surface morphology; (b) CL spectra of the MQW sample, including the whole region, the c-plane flat region and the semipolar sidewall region. (c) and (d) CL mapping images at wavelength of 407 and 457 nm, respectively. (e) PL spectra under different excitation power.

Fig. 3 Surface morphology and CL mapping of the LED sample. (a) and (b) Low and high magnification SEM observation on the surface morphology. (c) CL spectra of the LED sample, including the whole region, the c-plane flat region and the semipolar sidewall region. (d) and (e) CL mapping images at wavelength of 417 and 457 nm, respectively.

Fig. 4 LED EL photos under small injection current. The bright spots on surface corresponds to the V-pits locations for enhanced hole injection around them.

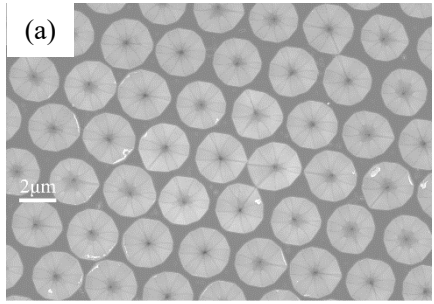 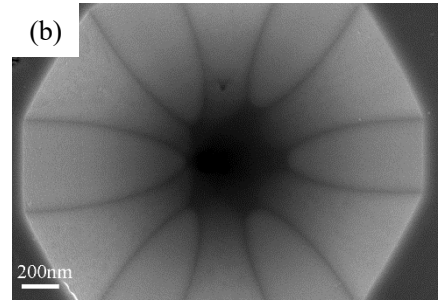

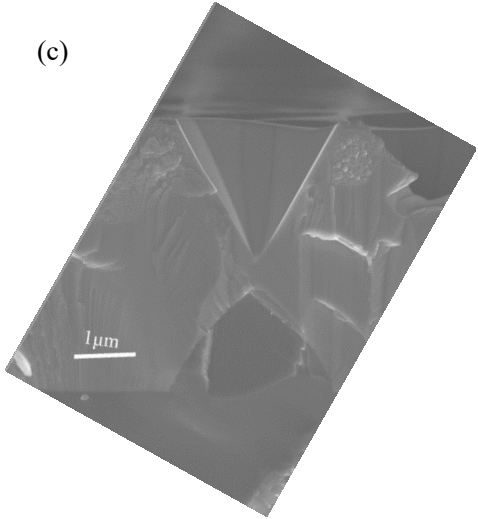

Fig.1 L. Wang *et al.*

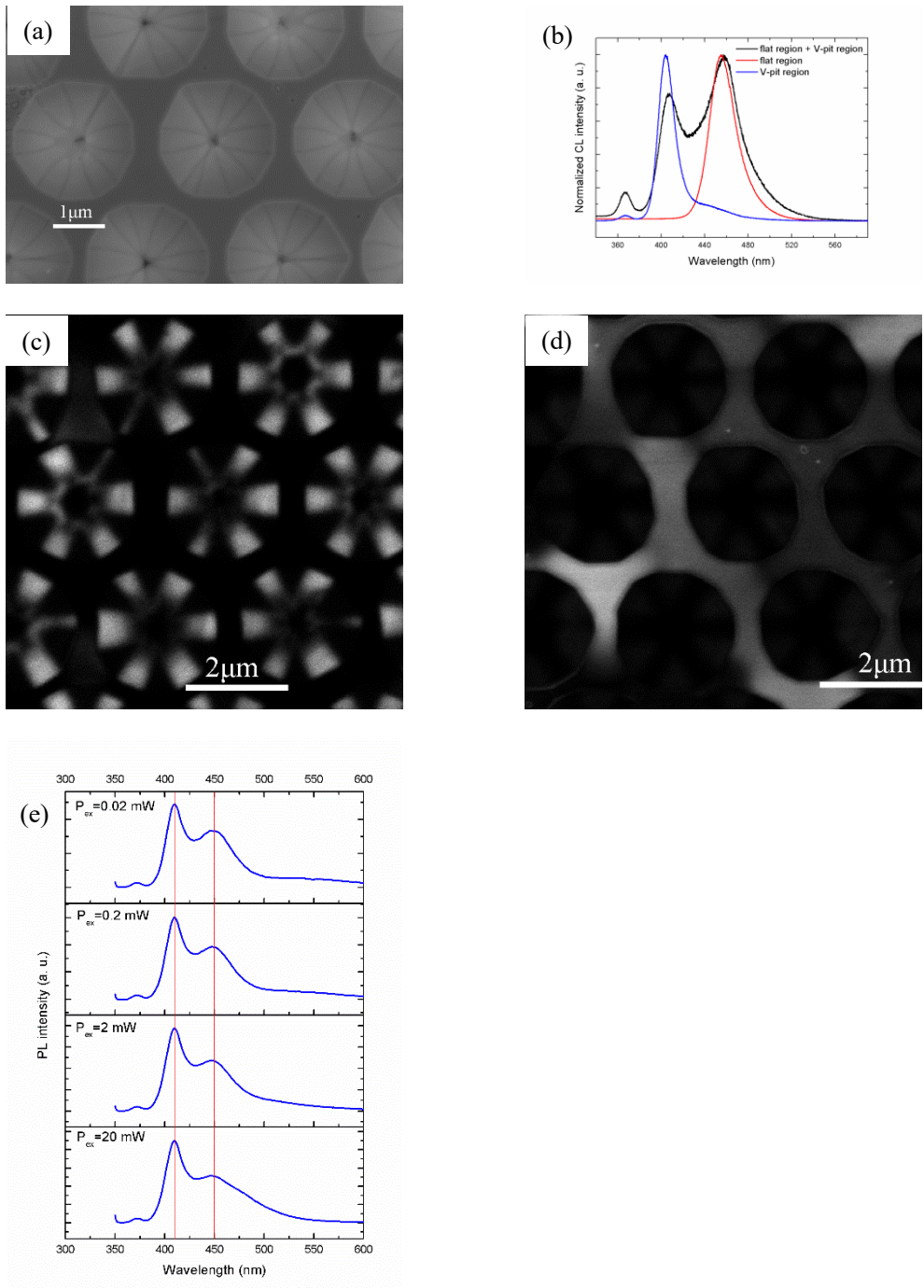

Fig.2 L. Wang *et al.*

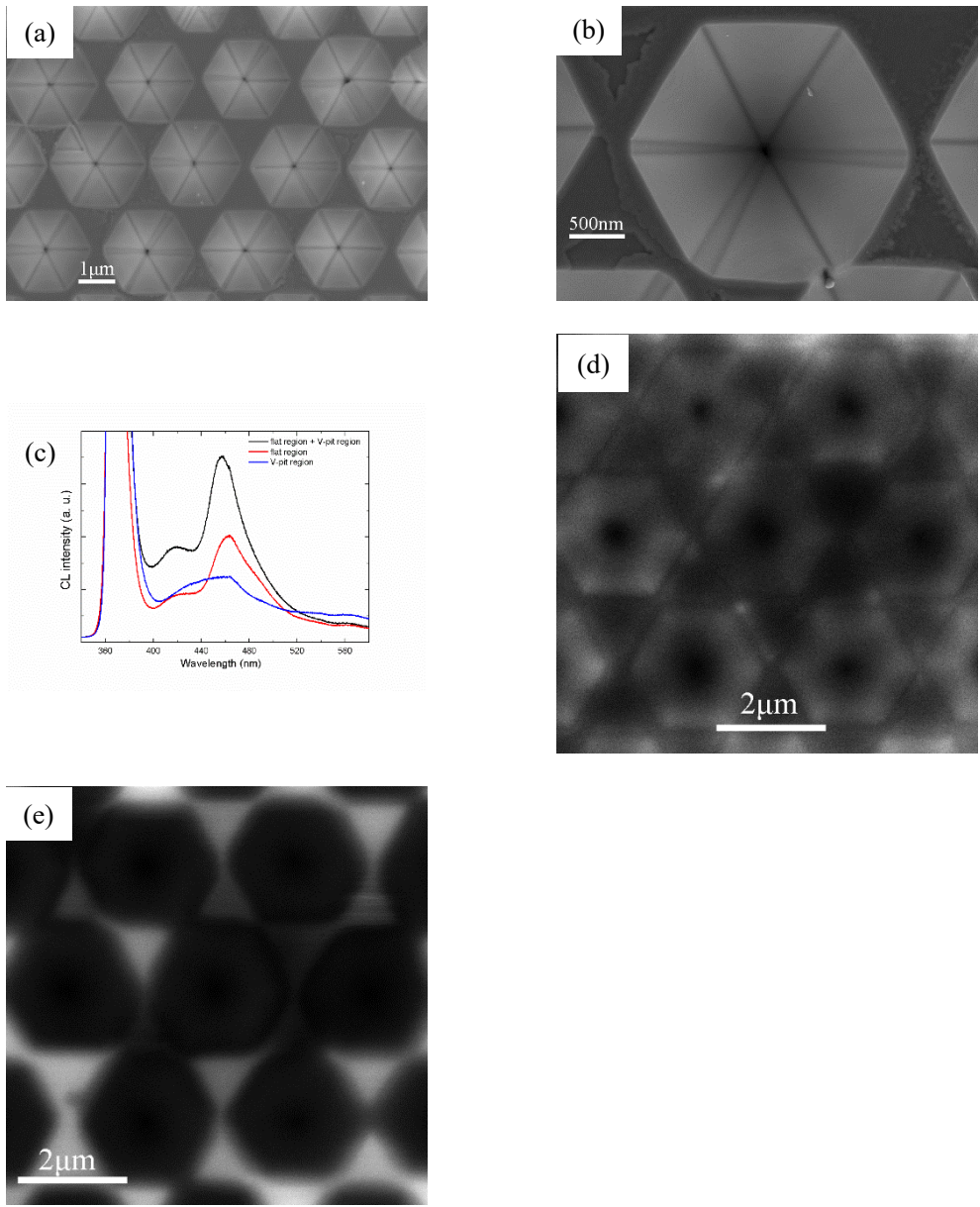

Fig.3 L. Wang *et al*.

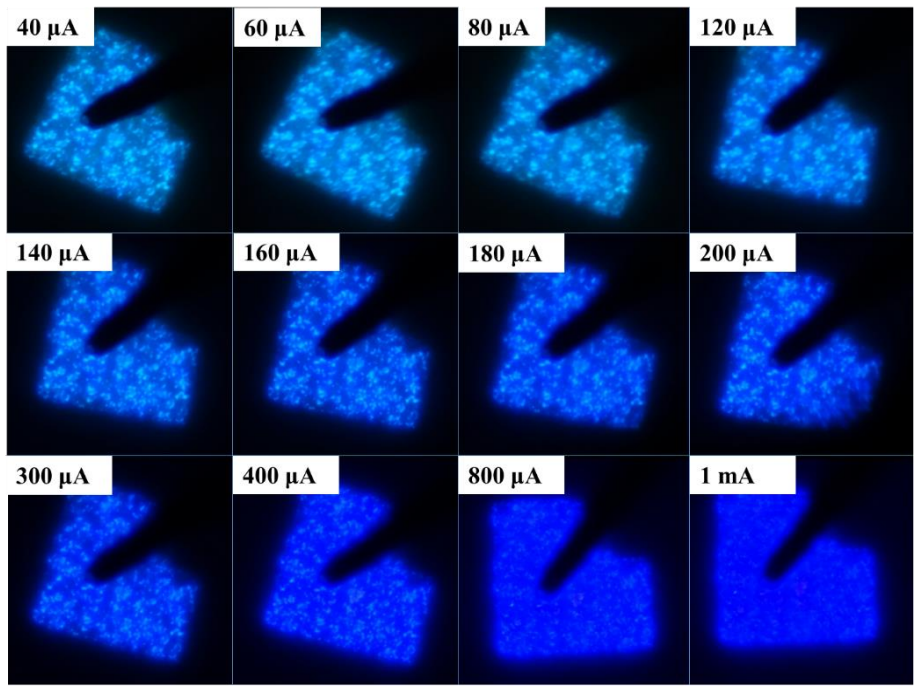

Fig.4 L. Wang *et al.*